\documentclass[twocolumn,showpacs,preprintnumbers,amsmath,amssymb,prb,aps]{revtex4}

\usepackage{epsfig} 
\usepackage{graphicx}
\usepackage{dcolumn}
\usepackage{bm}
\usepackage{epstopdf}
\usepackage{amsmath}
\begin{document}

\title{Evidence of spin lattice coupling in MnTiO$_{3}$: an x-ray diffraction study}
\author{R. K. Maurya$^1$}
\author{Navneet Singh$^1$}
\author{S. K. Pandey$^2$}
\author{R. Bindu$^1$}
\altaffiliation{Corresponding author: bindu@iitmandi.ac.in}
\affiliation{$^1$School of Basic Sciences, Indian Institute of Technology Mandi, Kamand, Himachal Pradesh- 175005, India\linebreak
$^2$School of Engineering, Indian Institute of Technology Mandi, Kamand, Himachal Pradesh- 175005, India}

\date{\today}
\begin{abstract}
Here we investigate the temperature evolution of the structural parameters of a potential magnetoelectric material, MnTiO$_{3}$. The experimental results reveal interesting temperature dependence of the $\emph{c}$/$\emph{a}$ ratio and the Mn-O bonds which can be divided into three regions. In region I (300 K to 200 K), the above parameters are seen to decrease with decrease in temperature due to thermal effect. In the region II (200 K to 95 K), the decrement in the structural parameters are reduced due the competing intra layer antiferromagnetic interaction setting in $\sim$ 200 K. The $\emph{c}$/$\emph{a}$ ratio are seen to display a minima around 140 K. Below 140 K, the short Mn-O bonds increase suggesting the onset of inter layer antiferromagnetic interaction $\sim$ 100 K. In region III (95 K to 23 K), the antiferromagnetic interaction is fully established. The behaviour of the calculated Mn-O bonds based on first principle calculations are in line with the experimental results. This study demonstrates the importance of spin lattice coupling in understanding the magnetic properties of the compound which is expected to be helpful in revealing the origin of magnetically induced ferroelectricity.

\end{abstract}


\maketitle

The physical properties of a solid is governed by the behaviour of the electrons in it. The properties become more complex and interesting if the motion of electron is decided by the interplay between charge, spin, lattice and orbital degrees of freedom. Such cross coupling is the marked feature of strongly correlated electron systems. One of the properties which has gained significant amount of attention from both application and fundamental physics point of view is the multiferroicity\cite{wang}. In these kinds of compounds, the magnetism can be controlled electrically and polarization can be controlled magnetically. This phenomenon is possible when there are  significant couplings among spin, lattice and electrical dipole moment. MnTiO$_{3}$ is one such system which exhibits magnetically induced ferroelectricity. The thin films of this compound show ferrotoroidicity\cite{Tokura_apl}. For a material to undergo ferrotoroidicity, it is necessary to break both the inversion and time reversal symmetries. In view of this, coupling of spin and lattice degrees of freedom is expected to be present. In this paper, we employ the x-ray diffraction (XRD) technique to study the spin lattice coupling in MnTiO$_{3}$ which is an important parameter to understand the multiferroic properties of this compound.

MnTiO$_{3}$, an ilmenite, stabilizes in hexagonal structure with centro symmetric R$\bar{3}$ space group \cite{shirane}.  Here the Mn$^{2+}$ ions are magnetic and Ti$^{4+}$ ions are non magnetic. This material undergoes paramagnetic (PM) to antiferromagnetic (AFM) transition at $\sim$ 64 K with a broad anomaly $\sim$ 100 K. The broad anomaly is attributed to the setting of two dimensional intra layer AFM interactions \cite{akimitsu,syono,Yamauchi,Jun,Mufti,Sticker}. Goodenough has worked out both the intra and inter layer superexchange interactions and have shown mainly two intra layer and three inter layer superexchange interactions \cite{Goodenough}.

MnTiO$_{3}$ also reveals linear magnetoelectric coupling similar to that observed in Cr$_{2}$O$_{3}$ \cite{seshadri}. At zero magnetic field, the spins are aligned along the $\emph{c}$-axis and no ferroelectric behaviour is observed. With the application of field, the material becomes ferroelectric and beyond a critical applied magnetic field of 6.5 T along the $\emph{c}$-axis, the spins tend to flop along the basal plane. At zero field, no anomaly is observed in the dielectric constant. Based on this fact, Mufti $\emph{et al.}$  \cite{Mufti} have pointed out that this material shows weak spin lattice coupling at the virgin state i.e. zero applied magnetic field. There are several factors which contribute to changes in the dielectric constant induced by magnetization. These include spin lattice coupling, electronic structure, orbital degrees of freedom, etc. Non observance of anomaly in the dielectric  constant due to spin lattice coupling indicates that the nature of the spin structure is such that no break in spatial symmetry is observed. It has been observed in CoCr$_{2}$O$_{4}$ that the spiral spin structure would lead to the break in the inversion symmetry to obtain magnetically induced ferroelectric order \cite{seshadri}.  Hence, the non observance of dielectric anomaly in the case of MnTiO$_{3}$ does not guaranty that there is insignificant spin lattice coupling. In addition, MnTiO$_{3}$ exhibits quasi two dimensional AFM. In such low dimensional system, the intra and inter layer superexchange interactions are expected to depend on the Mn-O, Mn-Mn bond lengths and Mn-O-Mn bond angles. Hence it is expected to show structural response across the region of intra and inter layer magnetic interactions.

In this work, we report the temperature evolution of the various structural parameters of MnTiO$_{3}$ across the magnetic phase transition of the compound by using XRD technique. Based on the behaviour of thermal expansion coefficient, the temperature dependent structural parameters are divided into 3 regions. In the region I, the behaviour of the structural parameters are due to simple thermal effect. As the material enters into the region II, the decrement in the lattice parameters is drastically reduced and is also reflected in the Mn-O and Mn-Mn bonds. A minima is also observed in the $\emph{c}$/$\emph{a}$ ratio. The experimental results were corroborated with the first principle calculations. These behaviours have been interpreted to the setting up of short range intra layer and inter layer AFM interactions $\sim$ 200 and $\sim$ 100 K, respectively. Apart from this, we also observe unusual behaviour in the Ti-O bonds across the region of magnetic phase transitions which still needs to be understood.

\begin{figure}
 \vspace{-1ex}
\includegraphics [scale=0.4, angle=0]{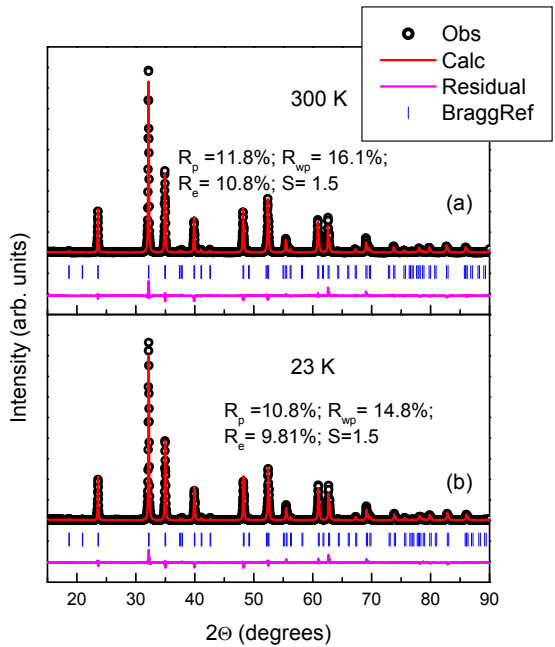}
 \vspace{-20ex}
\caption{Rietveld refinement of the XRD patterns of MnTiO$_{3}$ collected at (a) room temperature (300 K) and (b) 23 K. The open circle and solid line correspond to the observed and calculated patterns, respectively. The Bragg reflections are represented by ticks. R$_{p}$, R$_{wp}$, R$_{e}$ and S correspond to profile, weighted profile, expected weighted profile factors and the goodness of fit, respectively.}
\vspace{-2ex}
\end{figure}

The sample was prepared by conventional solid state route. The starting materials; MnCO$_{3}$ and TiO$_{2}$ were ground using mortar and pestle and sintered at 1200 $^{\circ}$C for 24 hours in air. The sample was characterized by using powder XRD and dc magnetization techniques. The results show complete solid solubility. The temperature dependent powder XRD experiments were performed using Smart lab 9 kW rotating anode x-ray diffractometer. The diffraction patterns were collected for 12 different temperatures to understand the temperature evolution of the structural parameters. The dc magnetization measurements were carried out using PPMS set up in the temperature range (300 K to 4 K) at applied field of 0.7 T.

The structures corresponding to the PM and AFM phases of MnTiO$_3$ have been relaxed by using state-of-the-art full potential linearized augmented plane-wave (FP-LAPW) method \cite{wien2k}. In these calculations we have fixed the lattice parameters to the experimental values of corresponding phases. The Muffin-Tin sphere radii were chosen to be 2.05, 1.76, and 1.59 a.u. for Mn, Ti, and O atoms, respectively. The recently developed PBESol \cite{pbesol} exchange-correlation functional have been used in the calculations. The convergence was achieved by considering 512 k points within the first Brillouin zone. The error bar for the energy convergence was set to be smaller than 10$^{-6}$ Ry/cell. The structure was considered to be relaxed when the total force became less than 1 mRy/a.u.

In Fig.1, we show the Rietveld refinement of the XRD patterns of MnTiO$_{3}$ collected at room temperature (RT) and 23 K. The RT data was indexed using R$\bar{3}$ space group with lattice parameters $a$= 5.1359(2) ${\AA}$, $c$=14.2782(5) ${\AA}$. The value of the lattice parameters are in line with other reports\cite{Mufti}. As the sample is cooled down to 23 K, there occurs no change in the number of reflections but shift in the peak positions is observed. Hence, this compound does not show any structural transition down to 23 K.

The magnetic susceptibility as a function of temperature, Fig.2a, exhibits broad peak $\sim$100 K and no anomaly is observed $\sim$T$_{N}$. Our experimental data are in line with the reported data \cite{Mufti}.The broad peak corresponds to 2 dimensional AFM interaction.The non appearance of sharp anomaly $\sim$T$_{N}$ has been attributed to the accidental cancellation of different interlayer interactions existing in this compound \cite{Jun}. Here, T$_{N}$ marks the transition from PM to 3 dimensional AFM phase. The value of T$_{N}$ is obtained by taking the first derivative of the susceptibility as a function of temperature which is $\sim$ 62.9 $\pm$ 1 K as shown in Fig.2a$'$. This value is in line with the reported value \cite{Mufti}.

\begin{figure}
\vspace{-2ex}
\includegraphics [scale=0.4, angle=0]{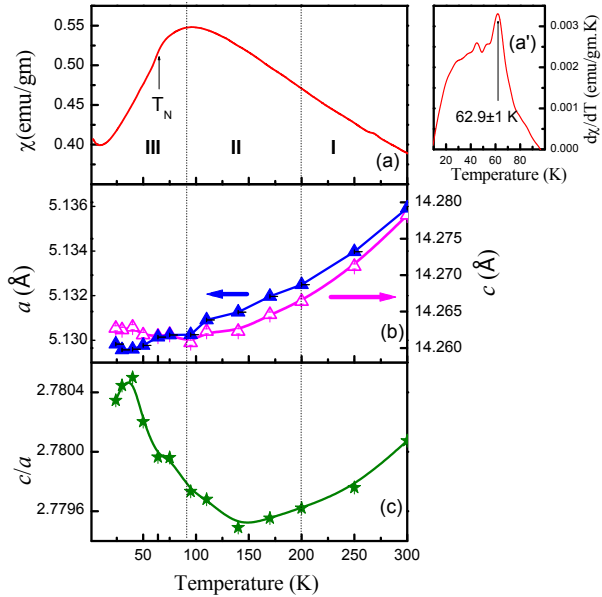}
\vspace{-15ex}
\caption{Temperature evolution of (a) DC magnetic susceptibility carried out at applied magnetic field 0.7 T; (a$'$) First derivative of susceptibility; (b) the lattice parameters and (c) $\emph{c}$/$\emph{a}$ ratio.}
\vspace{-2ex}
\end{figure}

To obtain the structural parameters across the region of magnetic phase transition, Rietveld profile refinement \cite{Rietveld,Rietveld1} were carried out at all the temperatures. In Figs.1a and b, we show typical Rietveld refinement of the XRD patterns. The goodness of fit (S) obtained at all the temperatures is close to 1.5. The temperature evolution of the lattice parameters $\emph{a}$ and $\emph{c}$ are shown in Fig.2b. The careful analysis of the data shows that the lattice parameters decrease with decrease in temperature but this decrement is not uniform throughout the temperature range of study. We observe that the decrement is significant until 200 K and later on until $\sim$ 95 K, this decrement decreases. On further reduce in temperature the lattice parameter $\emph{a}$ decreases marginally until 50 K and later on remains almost the same while $\emph{c}$ shows an increase below 95 K. Quantitatively, in the temperature range 300 K to 200 K, both the lattice parameters decrease by 0.07$\%$; in the range 200 K to 95 K, it decreases by 0.04$\%$; in the range below 95 K, the parameters shows marginal decrease until 50 K and later on it remains the same while the $\emph{c}$-parameter increases by 0.014$\%$. To obtain the thermal contribution, thermal expansion coefficient $\alpha$ was calculated for the lattice parameters in all the above mentioned temperature ranges. Here, we find that in the first temperature range, the value of $\alpha$ for $\emph{a}$ and $\emph{c}$ parameters are $\sim$ 6.7 x 10$^{-6}$ K$^{-1}$ and 8.3 x 10$^{-6}$ K$^{-1}$, respectively \cite{mineralogy}. In the second temperature range, it is 4.15 x 10$^{-6}$ K$^{-1}$ and 2.7 x 10$^{-6}$ K$^{-1}$ and in the third range, it is 2.03 x 10$^{-6}$ K$^{-1}$ and -1.898 x 10$^{-6}$ K$^{-1}$, respectively. Considering above facts, the behaviour of the structural parameters can be divided into 3 regions. The region I lies in the range 300 K to 200 K; region II in the range below 200 K to 95 K and region III is the range below 95 K to 23 K. The behaviour of the $\emph{c}$/$\emph{a}$ ratio is shown in Fig.2(c). Here, we observe that in the region I, the ratio decreases; in region II, it reaches a minima at around 140 K and in region III an increase is observed.

To understand this behaviour, it is important to look into the Mn-O, Ti-O, etc. bond distances. In the region I, the short Mn-O bonds (Mn-O(S)) decrease with decrease in temperature, Fig.3(a). In the region II, it is found to increase below 140 K, then decreases in the region III and later on remains almost the same. The long Mn-O bond (Mn-O(L)) lengths, Fig.3(b); are found to decrease until region II and in the region III, it increases from 75 K and below 50 K, it decreases. The Mn-Mn bonds are found to decrease in region I and a marginal increase is observed in the region II and in the region III, it remains almost the same.

\begin{figure}
\vspace{-2ex}
\includegraphics [scale=0.4, angle=0]{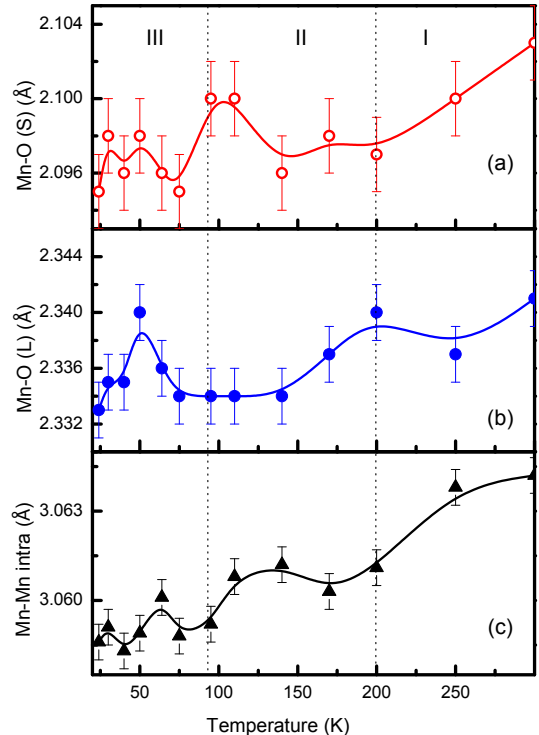}
\vspace{-10ex}
\caption{Temperature evolution of (a) Mn-O(S); (b) Mn-O(L) and (c) Mn-Mn intra bond lengths.}
\vspace{-2ex}
\end{figure}

In regions I and II, the short and long Ti-O bonds, Fig.4(a and b) are found to increase and decrease, respectively with decrease in temperature. On further reduce in temperature, both the bonds reveal opposite behaviour. Reduction in the Mn-Ti bonds, Fig.4(c) is observed in the region I and in the regions II and III, it remains almost the same.

The above results show interesting temperature evolution of various structural parameters across the magnetic phase transitions. However, the changes in bond lengths are small and in some temperature range these changes are within the error bar. At this stage it is important to note that x-ray diffraction technique provides extremely accurate value  for lattice parameters. The presence of error bar in fourth decimal place, as mentioned above, is an indication to this fact. However, this technique is not very accurate for determining the atomic positions of lighter elements especially the oxygen ions. This is because the x-rays interact with the electron cloud, so the contribution of the diffracted intensity for the heavier element is more as compared to the lighter elements. In the light of this fact the small changes seen in the bond lengths within the error bar may not be so reliable. In order to confirm the reliability of the experimental results across the magnetic phase transitions we have performed the DFT based first principles calculations by using Wien2k code \cite{} where the structure of the compound is relaxed in both PM and AFM phases. The computational details are given above. The structural parameters corresponding to both the phases obtained after relaxations are shown in Table 1. One can clearly see that the Mn-O(S) increases by +0.271 ${\AA}$ and Mn-O(L) bonds decreases by 0.0863 ${\AA}$ as the compounds enter from PM to the AFM phase. The Mn-Mn (intra) bond lengths are seen to increase by 0.0772 ${\AA}$. The change in the calculated Mn-O and Mn-Mn bond lengths across the magnetic phase transition is in line with the experimental data. However, the changes observed in the calculated bond lengths are more as compared to the experimental results. This is due to the fact that thermal contribution to the structural parameters is absent in the calculation which is competing with the magnetic contribution. Hence the experimentally observed changes in structural parameters are expected to be less than that observed in the calculations.

\begin{figure}
\vspace{-2ex}
\includegraphics [scale=0.4, angle=0]{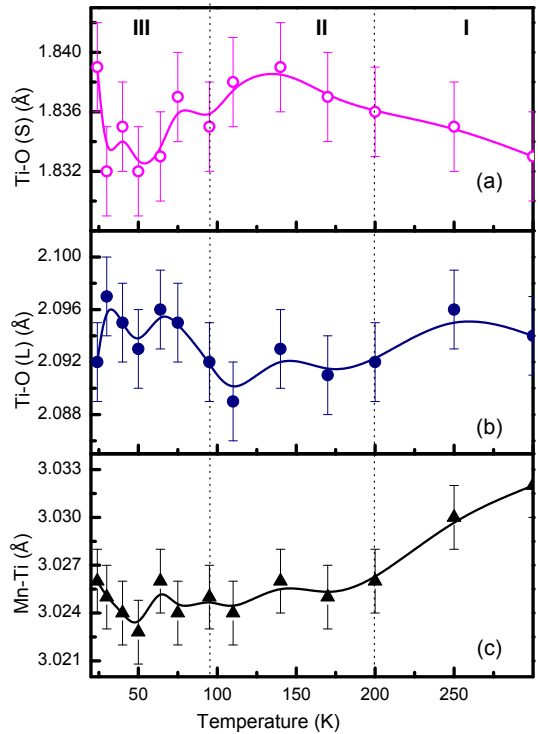}
\vspace{-10ex}
\caption{Temperature evolution of (a) Ti-O(S); (b)Ti-O (L) and (c) Mn-Ti bond lengths}
\vspace{-2ex}
\end{figure}

The above results clearly show the discernible changes in the structural parameters in the temperature range where various magnetic interactions set in. Thus the electronic and magnetic properties of this compound is expected to be closely related to the Mn-O and Mn-Mn bonds. This expectation appears to be a reality in the light of the fact that the magnetoelectric effect in this material arises due to Mn ions and the superexchange interaction depends strongly on the bond distances and bond angles. The behaviour of the structural parameters can be understood in the following way. In the region I, the structural parameters namely lattice parameters, Mn-O and Mn-Mn bonds show simple thermal contraction as the value of $\alpha$ are in line with the general ilmenite structures \cite{mineralogy}. As the material enters region II, $\sim$ 200 K, the reduction in the $\alpha$ value suggests the evolution of the short range intra layer AFM. To observe any effect on the diffraction pattern due to onset of intra layer AFM ordering, the coherence length should be atleast some hundreds of angstroms. So the signature of the short range magnetic cluster in the diffraction pattern suggests the spatial extent of the magnetic phase is sufficient enough to be reflected in the structural parameters.

\vspace{2ex}
\begin{tabular}{|c|c|c|c|c|}
  \hline
   & Paramagnetic & Antiferromagnetic\\
     \hline
  $Mn-O(s)$ (${\AA}$) & 1.9801 & 2.2511 \\
    \hline
  $Mn-O(L)$ (${\AA}$) & 2.1743 & 2.0880 \\
    \hline
  $Mn-Mn (intra)$ (${\AA}$) & 2.9663 & 3.0435 \\
    \hline

  \end{tabular}

\vspace{2ex}

Table 1: Bond lengths obtained using first principle calculations in the paramagnetic and antiferromagentic phases.

\vspace{2ex}

In MnTiO$_{3}$, Goodenough $\emph{et al}$ \cite{Goodenough} have shown that there are five superexchange interactions in this material. Out of these two are intra layer and the rest are inter layer ones. The sign of the exchange interaction depends on the number of electrons in the orbitals which are interacting. For half-filled interacting orbitals superexchange interaction is antiferromagnetic. When there is more than one electron per interacting orbital, the interaction is ferromagnetic. Following the Goodenough Kannamori rules\cite{goodenough_book}, the intra layer AFM interaction occurs through the overlap of half-filled orbitals at the Mn sites. Hence in the region II, $\sim$ 200 K, there occurs onset of the magnetic interaction within the layer and is depicted as decrease in the Mn-O (L) and Mn-Mn (intra) bond lengths above 140 K. It is interesting to note that the increase in the $\emph{c}$/$\emph{a}$ ratio below 140 K and also increase in the Mn-O(S) bonds suggest the onset of the weak interlayer super superexchange interaction between the cations through the oxygen-oxygen ions. The increase in the Mn-O(S) bond length leads to the decrease in the separation between the O-O bonds lying along the $\emph{c}$-axis and hence enhancement in the super superexchange interactions between the layers. So the critical temperature for setting up of the intra and inter layer AFM interactions occurs in the region II. So far there are no reports regarding this aspect.

In the region III, for temperatures below 50 K, contribution due to thermal effect and AFM interaction to the $\emph{a}$ parameter is almost the same. Hence almost no change in the value of $\emph{a}$ is observed. While for the $\emph{c}$ parameter the magnetic interaction dominates thereby leading to its increment. The behaviour of the lattice parameters are more evident in the $\emph{c}$/$\emph{a}$ ratio plot shown in Fig.2(c). It is also observed that the Mn-O(S) bonds initially decrease and remains the almost the same until the lowest collected temperature while the Mn-O(L) bonds shows an increase at around 50 K and later on remains almost unchanged.

Apart from the Mn-O bonds, the Ti-O bonds also reveal significant changes across the region of intra and inter layer AFM interaction. However, it is still not clear about the role of Ti in magnetism and magnetically induced ferroelectricity exhibited by this system. We hope that careful observation of the local structural studies will provide clue towards answering this question.

In summary, we have studied the temperature evolution of the structural parameters of MnTiO$_{3}$ using x-ray diffraction technique. The temperature behaviour of the structural parameters was divided into three regions; (i) 300 to 200 K, (ii) 200 K to 95 K and (iii) 95 K to 23K. In region I, the decrease in the $\emph{c}$/$\emph{a}$ ratio and the Mn-O bonds with decrease in temperature is attributed to thermal effect. In region II, $\sim$ 200 K, decrement in the above parameters is altered as a result of the competition between the intra layer AFM interaction and the thermal effect. Around 140 K, a minima is observed in the $\emph{c}$/$\emph{a}$ ratio. The short Mn-O bonds shows an increase below this temperature suggesting the onset of intra layer AFM interaction. The bond lengths obtained using first principle calculations is in consonance with the experimental results. From the present results, it appears that spin lattice coupling plays significant role in stabilising magnetically induced ferroelectricity in this system. X-ray diffraction studies in the presence of magnetic field is expected to be helpful in addressing the origin of magnetically induced ferroelectricity in this compound.

\end{document}